 \documentclass[authoryear,final,5p,times,twocolumn]{elsarticle}

 \usepackage{graphicx}
\usepackage{amssymb}
\def\W18b{WASP-18b}

\def\deg{$^\circ$}

\usepackage{color}

\usepackage{aas_macros} %%% inclut les abreviation ADS dans la biblio

\journal{Icarus}

\begin{document}

\begin{frontmatter}

%% Title, authors and addresses

%% use the tnoteref command within \title for footnotes;
%% use the tnotetext command for the associated footnote;
%% use the fnref command within \author or \address for footnotes;
%% use the fntext command for the associated footnote;
%% use the corref command within \author for corresponding author footnotes;
%% use the cortext command for the associated footnote;
%% use the ead command for the email address,
%% and the form \ead[url] for the home page:
%%
%% \title{Title\tnoteref{label1}}
%% \tnotetext[label1]{}
%% \author{Name\corref{cor1}\fnref{label2}}
%% \ead{email address}
%% \ead[url]{home page}
%% \fntext[label2]{}
%% \cortext[cor1]{}
%% \address{Address\fnref{label3}}
%% \fntext[label3]{}

%\title{An atmospheric model for the hot transiting exoplanet WASP-18b}
\title{On the heat redistribution of the hot transiting exoplanet WASP-18b}

%% use optional labels to link authors explicitly to addresses:
%% \author[label1,label2]{<author name>}
%% \address[label1]{<address>}
%% \address[label2]{<address>}

\author[label1,label2]{N. Iro\fnref{foot1}}
\ead{nicolas.iro@uni-hamburg.de}%{iro@astro.keele.ac.uk}
\author[label1]{P.~F.~L. Maxted}
\ead{p.maxted@keele.ac.uk}%{pflm@astro.keele.ac.uk}
\address[label1]{Astrophysics Group, Keele University, Keele, Staffordshire, ST5 5BG, UK}
\address[label2]{LESIA, Observatoire de Paris, 5 Place Jules Janssen, F-92195 Meudon, France}
\fntext[foot1]{Now at: Theoretical Meteorology, Meteorological Institute - University of Hamburg, Grindelberg 5, 20144 Hamburg, Germany}
\begin{abstract}
%% Text of abstract
The energy deposition and redistribution in hot Jupiter atmospheres is not well understood currently, but is a major factor for their evolution and survival. We present a time dependent radiative transfer model for the atmosphere of WASP-18b which is a massive (10 $M_{\rm Jup}$) hot Jupiter ($T_{\rm eq}~\sim~2400$~K) exoplanet orbiting an F6V star with an orbital period of only 0.94 days.

Our model includes a simplified parametrisation of the day-to-night energy redistribution by a modulation of the stellar heating mimicking a solid body rotation of the atmosphere.
We present the cases with either no rotation at all with respect to the synchronously rotating reference frame or a fast differential rotation.
 
The results of the model are compared to previous observations of secondary eclipses of \citet{Nymeyer10} with the {\it Spitzer} Space Telescope.
%new results from a Spitzer campaign to observe lightcurves at 3.6-micron and 4.5-micron.
%The observed phase curve flux variation -- in particular the extremely dark night side of the planet -- 
Their observed planetary flux suggests that the efficiency of heat distribution from the day-side to the night-side of the planet is extremely inefficient. 
%The amplitude, phase and shape of the phase variation in WASP-18b were also measured.
%This enables us to put strong constraints on the efficiency of heat redistribution in this planetary system and to determine the pressure level at which the heat distribution occurs.
%We confirm 
Our results are consistent with the fact that such large day-side fluxes can be obtained only if there is no rotation of the atmosphere.

Additionally, we infer light curves of the planet for a full orbit in the two Warm \emph{Spitzer} bandpassses for the two cases of rotation and discuss the observational differences.

\end{abstract}

\begin{keyword}
%% keywords here, in the form: keyword \sep keyword
Atmospheres, structure \sep
Extrasolar planets \sep
Radiative transfer

%% MSC codes here, in the form: \MSC code \sep code
%% or \MSC[2008] code \sep code (2000 is the default)

\end{keyword}

\end{frontmatter}

%%
%% Start line numbering here if you want
%%
% \linenumbers

%% main text
\section{Introduction}
\label{Intro}

\W18b\ \citep{Hellier09} is a massive planet orbiting its F6V parent star.
The high mass and short period of WASP-18b make it a useful test-case for exploring the tidal interaction between planets and their host stars \citep{Brown11, Matsumura10}.
%(Brown et al., 2011MNRAS.415..605B, Matsumura et al., 2010ApJ...725.1995M). 
Accurate parameters for WASP-18b from high quality lightcurves have been presented by 
\citet{Southworth09}.
%Southworth et al. (2009ApJ...707..167S). 
The observed Rossiter-McLaughlin effect for WASP-18 suggests that the  orbital axis of the planet is aligned with the rotation axis of the star
\citet{Triaud10}.
%(Triaud et al., 2010A&A...524A..25T).

\citet{Nymeyer10} studied two secondary eclipses of \W18b\ obtained with the four IRAC channels of the {\it Spitzer} Space Telescope.
As a consequence of the high mass of the planet, the signal we measure originates from deeper in the atmosphere (higher pressure) than for a lower mass planet.
These author infer indeed a contribution function peaking at 10~bars for 3.6~$\mu$m in a case without thermal inversion.
For a lower mass planet such as HD189733b ($M_p \sim 1.14 M_{\rm Jup} $), the same bandpass sounds an atmospheric pressure of around 0.3~bar \citep{Knutson09}.

The analysis of \citet{Nymeyer10} suggests that the planet has a very low albedo as well as an inefficient heat redistribution.
They presented a parametrised atmospheric model following the method of \citet{Madhu09}
that marginally favors a thermal structure presenting an inversion, as expected for planets receiving such high level of irradiation \citep{Burrows06, Fortney08}, even though a non inverted model fits the observations within the error bars.

This planet is among the ones we should be able to derive strong constraints on its heat redistribution or circulation pattern.
Indeed, in principle, the {\it Spitzer} Space Telescope could be used to 
construct a day to night map of heat redistribution following the method from \citet{Knutson07}.
These authors observed HD189733b during a full orbit in order to get the phase resolved brightness temperature distribution.
This kind of study for this massive planet would enable us to test atmospheric models under a different regime by constraining a complementary pressure range.

In Section~\ref{Model}, we describe the radiative transfer model used in this work and the method used to parametrize the heat redistribution.
In Section~\ref{Results}, we present and discuss the results including the thermal structure and spectra as well as phase curves of the planet.

\section{Atmospheric model}
\label{Model}

\subsection{Radiative transfer}

We use line-by-line radiative transfer calculations from 0.3 to 25~$\mu$m, described in detail in  \citet{Iro05}.

In their modelling of \W18b , \citet{Nymeyer10} applied the method of \citet{Madhu09} where 
$\sim 10^7$ models are computed with six parameters describing the three layers of the $P$-$T$ profile and four parameters controlling
the departure from chemical equilibrium of the four major species (H$_2$O, CO, CH$_4$ and CO$_2$).
A $\chi^2$ minimization is performed between the numerous models against the four observational data points.
Their goal is to eliminate ranges of parameters that are too far from the observational constraints.

On the contrary, in our approach we calculate the chemical abundances by assuming that the atmosphere is in thermochemical equilibrium with a solar abundance of the elements.
In a more complete model, NLTE processes should be taken into account.
The model has 56 pressure levels from about 3.5~kbars to 10$^{-3}$~mbar.
By comparing the radiative lapse rate obtained with the adiabatic one, we adjust for the convective zone.

As opacity sources, we include the main molecular constituents: H$_2$O,
CO, CH$_4$, Na, K and TiO; Collision Induced Absorption by H$_2$ and He;
Rayleigh diffusion and H$^-$ bound-free and H$_2^-$ free-free.
We did not include CO$_2$ but in thermochemical equilibrium, its abundance is low and we expect that it does not impact on the radiative budget.
Moreover, the current model does also not account for clouds.

We can calculate the evolution of the temperature
profile in an atmosphere under hydrostatic equilibrium by using a time-marching algorithm, that allows us to include time-varying conditions.
By using this approach, different from \citet{Madhu09}, we are able to introduce a parametrized prescription of a rotation of the atmosphere.
This can be related to an estimate of the wind speed and, on the first order, the efficiency of the heat redistribution from day side to night side.

\subsection{Time-dependent calculations}

After convergence to a static solution corresponding to conditions where the planet is under globally-averaged insolation, we introduced a variation with time to the heating due to the star insolation.
We set as outer boundary condition a time-dependent incoming stellar
flux (more details can be found in \citealp{Iro10}):
\begin{equation}
F_\star^{\rm inc}(t,\lambda,\theta)= \left( \frac{R_\star}{a(t)}\right)^2
\cos \left[\theta(t)\right] F_\star^0(\lambda)
\end{equation}
At each time-step, we calculate the star radius $R_\star$, the
star-to-planet distance $a(t)$, $\theta$ the planetary longitude of the point that
is substellar at $t=0$ and $F_\star^0$ the flux emerging from the star for
each wavelength $\lambda$.
As a consequence, each parcel of atmosphere receive a stellar flux as if the global atmosphere were in rotation at a velocity set up by $\theta(t)$. The planet itself is assumed to be in synchronous rotation.

We performed calculations in two cases: one where the incoming flux depends only on the longitude $\theta$, constant with time in that case (flux maximum at the substellar point, nill for the night side); and one where we introduce a solid body rotation ($\theta$ varies with time) corresponding to a velocity of about 6.5 km~s$^{-1}$ at 1 bar.
In \citet{ShowmanPolvani}, superrotating jets at the equator with speed of the order of  the km~s$^{-1}$ can be maintained by angular momentum pumping from regions outside the jet to within the jet ('up-gradient eddy flux' proposed by \citealp{Cho08}).
Several eddy transport or planetary scale waves can be responsible for such transfer.
A possibility proposed by these authors is Rossby waves.
It should be noted that other mechanisms can generate this equatorial super rotation (see \citealp{Cho08b}).
Additionally, in the radiative-hydrodynamical simulations by \citet{Dobbs12}, zonal winds develop with velocity from $\sim 2$~km~s$^{-1}$ at 0.1bar to $\sim 4$ at $10^{-3}$ bars, in the case of HD189733b.
We then choose 6.5 km~s$^{-1}$ as an exploratory value in order to asses an upper limit of the wind velocity.

\subsection{Input parameters}

The planet, star and orbit parameters are taken from \citet{Southworth09}.
They are summarized in Table~\ref{tab:param}.

\begin{table}[htb]
  \begin{center}
  \caption{Planetary, stellar and orbital parameters used in the model \citep[from][]{Southworth09}.\label{tab:param}}
	  \begin{tabular}{llc}\hline \hline
      &Parameter& Value   \\
\hline
\vspace{2pt}
Orbit\\
	& $a$	& 0.02 AU\\
	& $P$	& 0.94 d\\
	& $e$ 	& 0.03 \\ 
\hline
\vspace{2pt}
Planet\\
	&$R_p$ & 1.165 $R_{\rm Jup}$\\
	&$M_p$ & 10.4  $M_{\rm Jup}$\\
	& $T_{\rm eq}$ & 2400 K\\
	& $T_{\rm eq}$(day) & 3100 K\\
\hline
\vspace{2pt}
Star\\
	& Type &F6V\\
	& $T_{\rm eff}$ & 6400 K\\
	& $R_\star$& 1.23 $R_\odot$\\
	    \hline \hline
	  \end{tabular}
  \end{center}
\end{table}

The equilibrium temperature is defined as:
$T_{\rm eq} = \left(1-A\right)^{1/4} \times f^{1/4} \times T_\star \times \left( R_\star/a \right)^{1/2}$ where $A$ is the planet albedo; f the heat redistribution factor of the planet's atmosphere; $T_\star$ the star's effective temperature; $R_\star$ the stellar radius and $a$ the semi-major axis of the orbit.
Using the parameter of the system from Table~\ref{tab:param}, this gives in the case of $A = 0$: $T_{\rm eq} =$~2400~K considering a heating averaged over the whole planet ($f=1/4$) and 2900~K considering an average over the day-side hemisphere ($f = 1/2$).
This makes WASP-18b a \emph{very} hot Jupiter.
For such highly irradiated planets, we expect a temperature inversion on the day-side \citep{Burrows06, Fortney08}.

The stellar flux has been computed by scaling an F-type star synthetic template from Castelli and Kurucz \citep{CastelliKurucz03}\footnote{See {\tt ftp://ftp.stsci.edu/cdbs/grid/ck04models}} for an effective temperature of 6400K.

\section{Results}
\label{Results}

\begin{figure}[htb]
  \begin{center}
    \includegraphics[width=0.9\columnwidth]{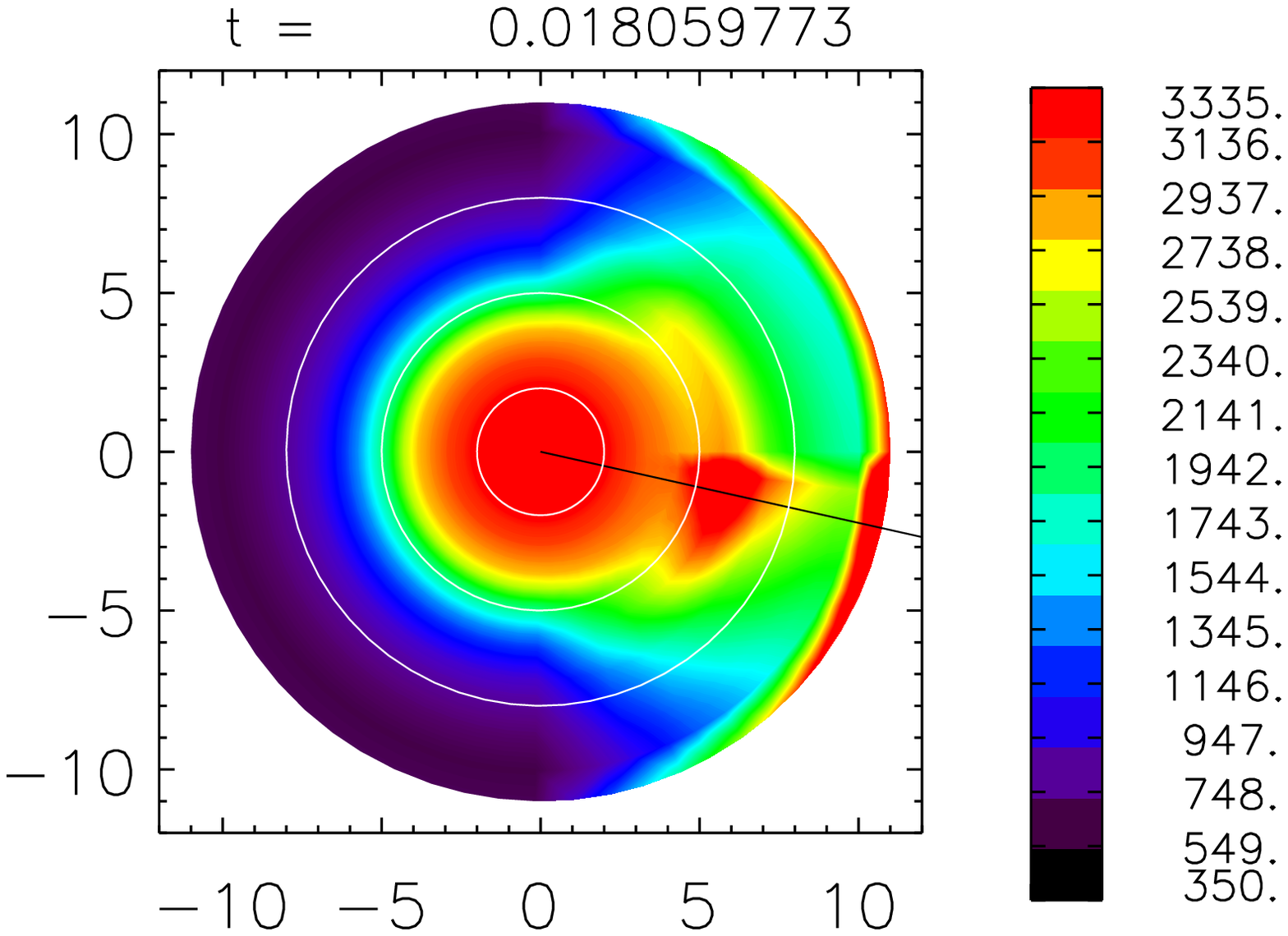}
    \includegraphics[width=0.9\columnwidth]{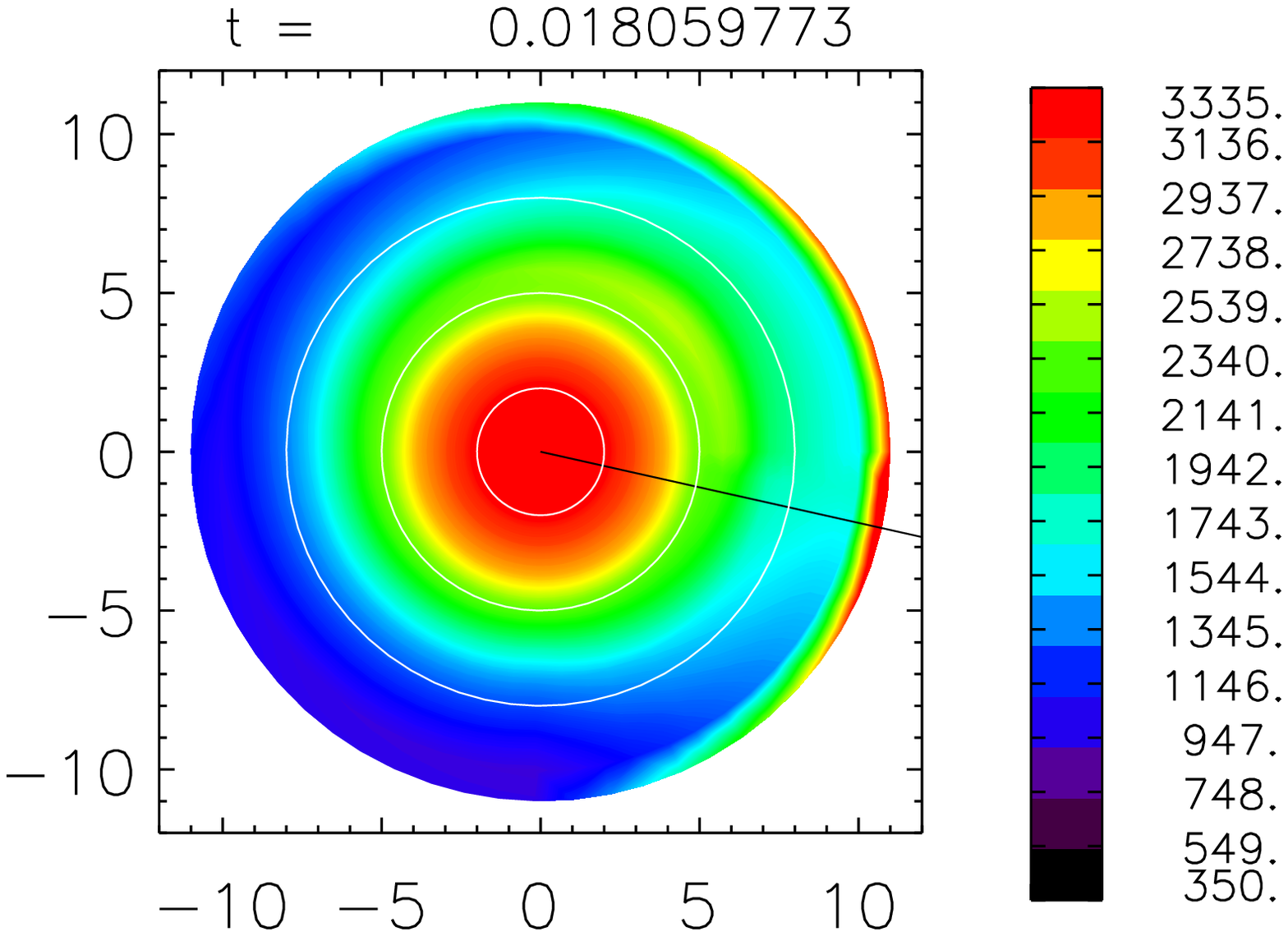}
    \caption{Equatorial cuts of the planet thermal structure\label{fig:maps} for the two cases considered. Upper panel: no rotation, lower panel: fast ($\sim$ 6.5km~s$^{-1}$ at 1 bar) rotation.
The colors represents the temperature in Kelvin.
The black line shows the direction of the star (substellar point). The pressure levels are shown in logarithmic scale with 10$^{-3}$ mbar at the top of the atmosphere and one unit represent one order of magnitude.
For quick review, three white circles are drawn at respectively 1 mbar, 1 bar and 10$^3$ bar from exterior to center.}
  \end{center}
\end{figure}

The temperature structure of the planet is illustrated by equatorial cuts at the equator in Figure~\ref{fig:maps}.
The effect of the rotation is obvious:
the more efficient heat redistribution in the longitudinal direction leads to a more mixed atmosphere where the temperature contrasts are dimmer.
The night-side gets warmer and the night-side cooler.
In the no-rotation case, day-to-night temperature contrasts go from 2,000~K to 3,000~K depending on the pressure level.
With a fast rotation, temperature contrasts drop to $\sim$~800K for pressures lower than 1~bar.

\begin{figure}[htb]
  \begin{center}
    \includegraphics[width=0.9\columnwidth]{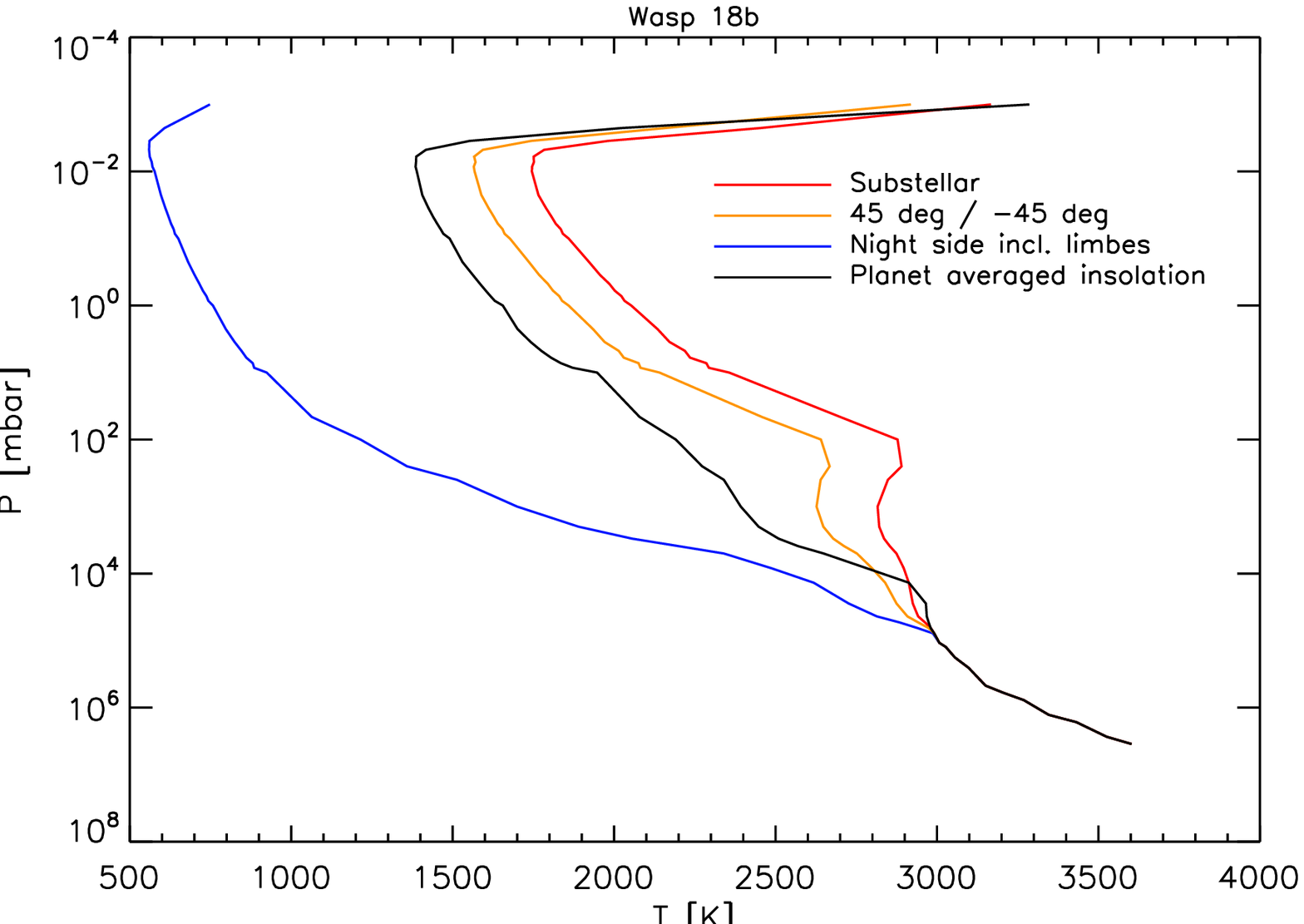}
    \includegraphics[width=0.9\columnwidth]{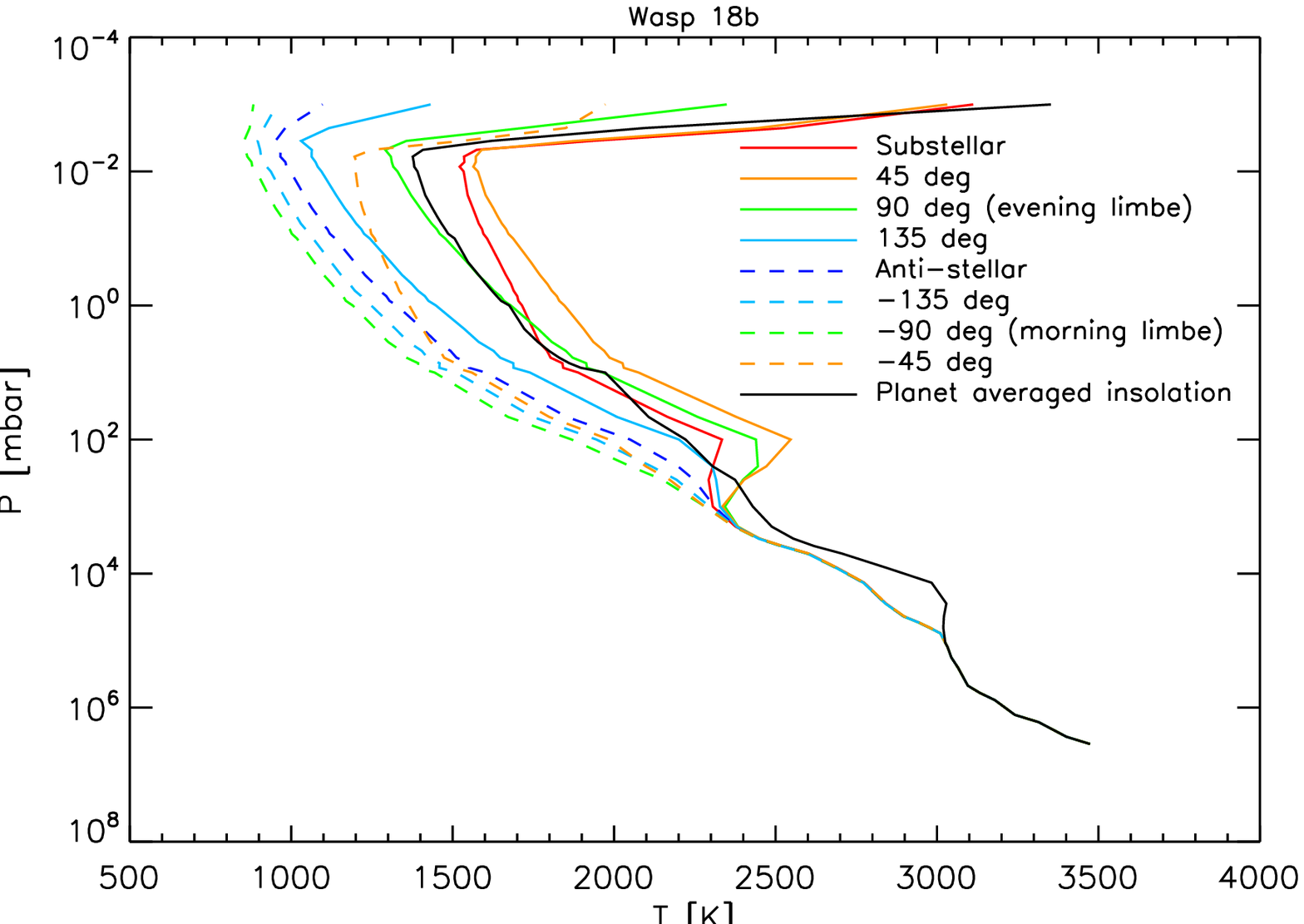}
    \caption{Thermal profiles for selected longitudes\label{fig:profs} for the two cases considered. Upper panel: no rotation, lower panel: fast rotation.
}
  \end{center}
\end{figure}

Figure~\ref{fig:profs} shows thermal profiles at selected longitudes.
These can be compared to Figure~13 of \citet{Nymeyer10}.
The deep ($>$ 1 bar) temperatures we calculate are lower than the one obtained by \citet{Nymeyer10}. This is due to the fact that in our approach, we initially have the temperature averaged for the whole planet and not the day-side.
Consequently, the net heat flux reaching the deep levels of their 1D model is substantially higher than for the current work.
In our model we have an inversion in the 1-0.1~bar pressure range.
In the no-rotation case, the temperature is maximum at the substellar point peaking at 2900~K.
When rotation is included, the thermal structure is more complex.
An inversion is still generated on the day-side but the hottest profile is for a longitude of 45\deg\ with a peak at $\sim$~2600~K and even the evening limb (90\deg ) is hotter than the substellar point.
These considerations are particularly interesting since it is the expected level that \emph{Spitzer} observations are probing.

\begin{figure}[htb]
  \begin{center}
    \includegraphics[width=0.9\columnwidth]{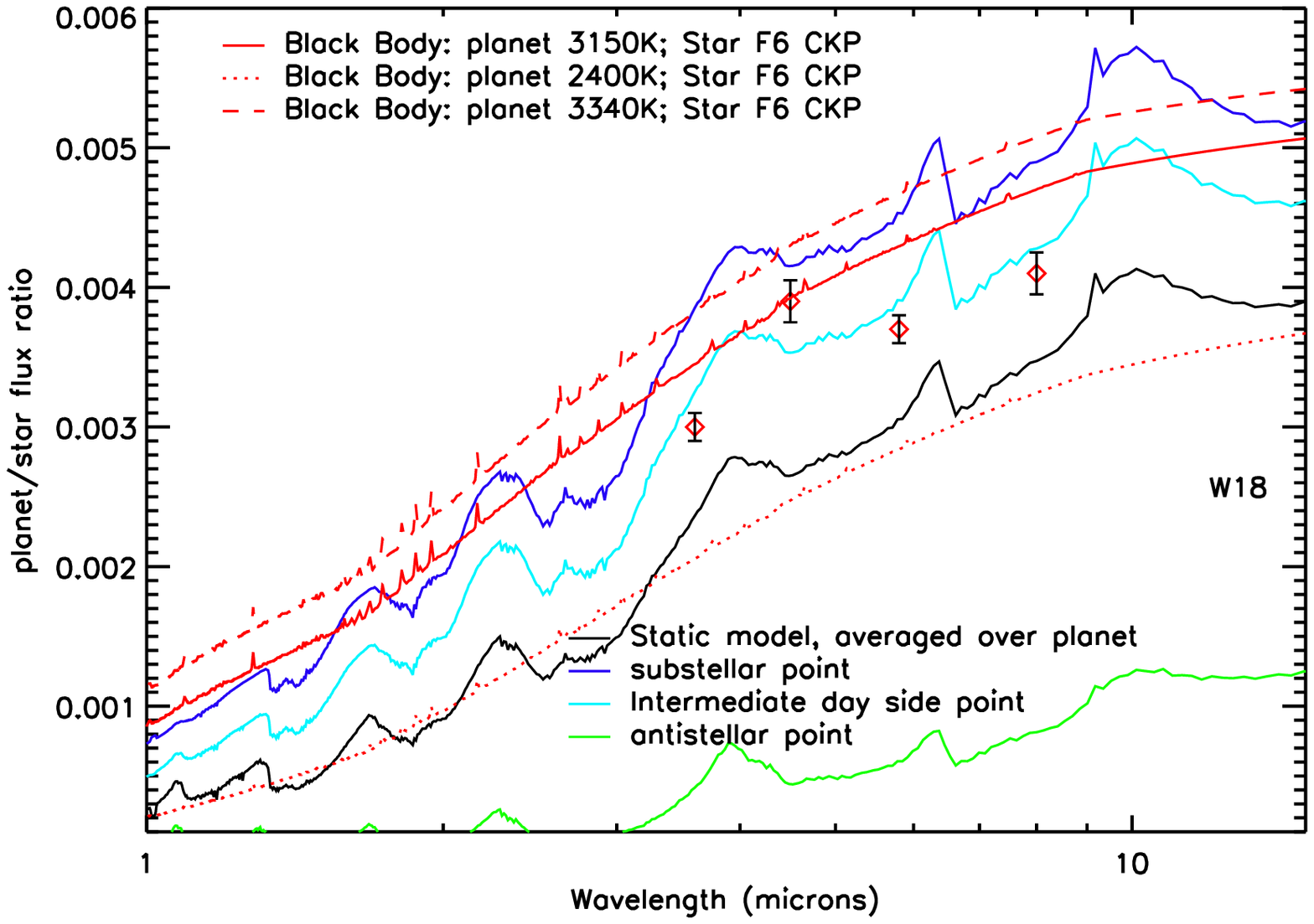}
    \includegraphics[width=0.9\columnwidth]{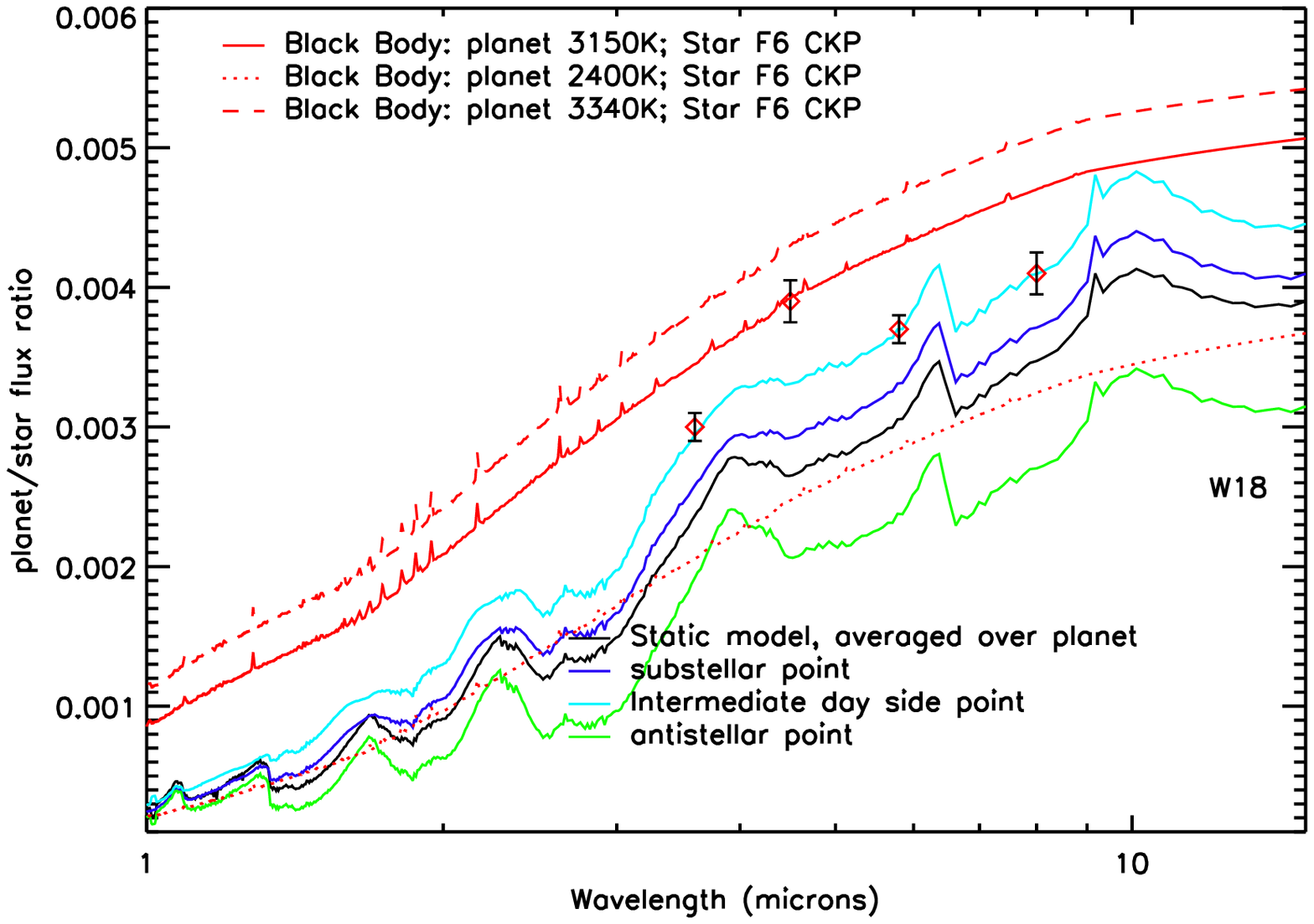}
    \caption{Flux ratio of the planet over the star\label{fig:fluxratio} as a function of wavelength for the two cases considered. Upper panel: no rotation, lower panel: fast rotation.
Several curves using a black-body planet are shown, with temperatures  representative of the night side and for a planet-averaged stellar heating.}
  \end{center}
\end{figure}

Figure~\ref{fig:fluxratio} represents the planet-to-star flux ratio as a function of wavelength for several positions --
at the substellar point, at an intermediate longitude (representative of an average over day-side) at the night side and also for the planet in a planet-averaged stellar heating condition.

Comparing the no-rotation with rotation cases, we can see the effect of decreased contrast (with rotation) between night and day.
The cooler day-side exhibits indeed a lower flux whereas the warmer night side has an increased flux compared to the no-rotation case.

The fast rotating case experiences fluxes too low on the day side to reproduce the values observed by \citet{Nymeyer10}, whereas they are very well fitted in the non rotating case.
This is consistent with these authors conclusion that WASP-18b should have very little redistribution of heat from the day side to the night side.

A trend seems to emerge: new observational constraints confirm that very highly irradiated planets have very poor day to night energy redistribution, as another example is given by WASP-12b whose phase curve has been measured by \citet{Cowan11}.
As stated by these authors, several factors contribute to an ineffective redistribution for very hot planets:
magnetic drag, very efficient at such large temperatures \citep{Perna10}, as emphasized by the GCM simulations by \citet{Rauscher12} where magnetic drag substantially slow down winds in HD209458b for a planetary magnetic field of at least 3G (even if thermal ionization could be insufficient in nominal cases, substantial ionization by stellar radiation is possible, see \citealp{Koskinen10});
the radiative timescale decreasing with temperature a lot faster than the advective timescale;
and by the peaks of thermal emission and stellar heating being close, a weak greenhouse effect.
We refer to the above articles for a more detailed study.
One can also mention that heat redistribution can have non-trivial mechanisms by waves \citep{Watkins10} and by instabilities \cite{Policht12}

\begin{figure}[htb]
  \begin{center}
    \includegraphics[width=0.9\columnwidth]{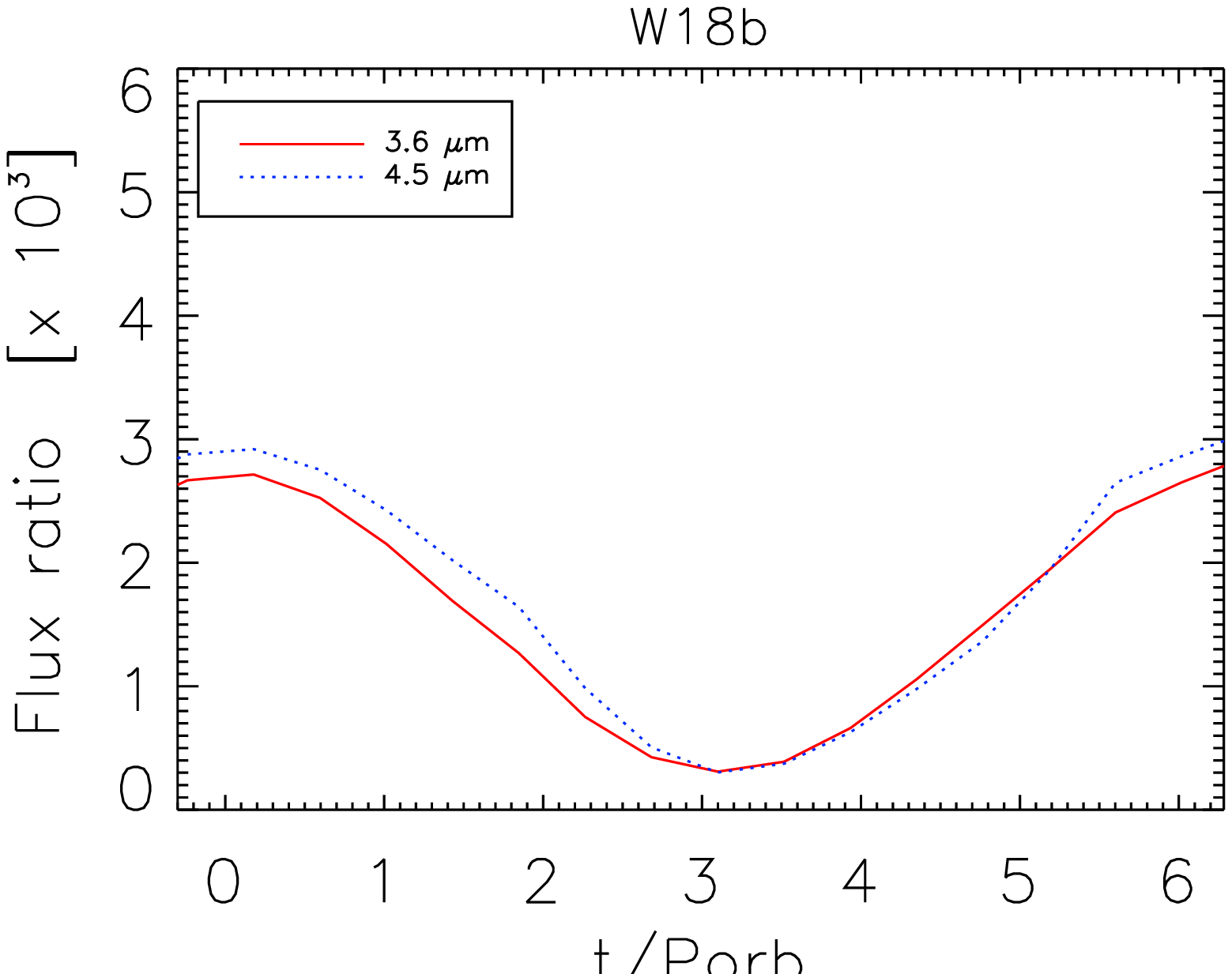}
    \includegraphics[width=0.9\columnwidth]{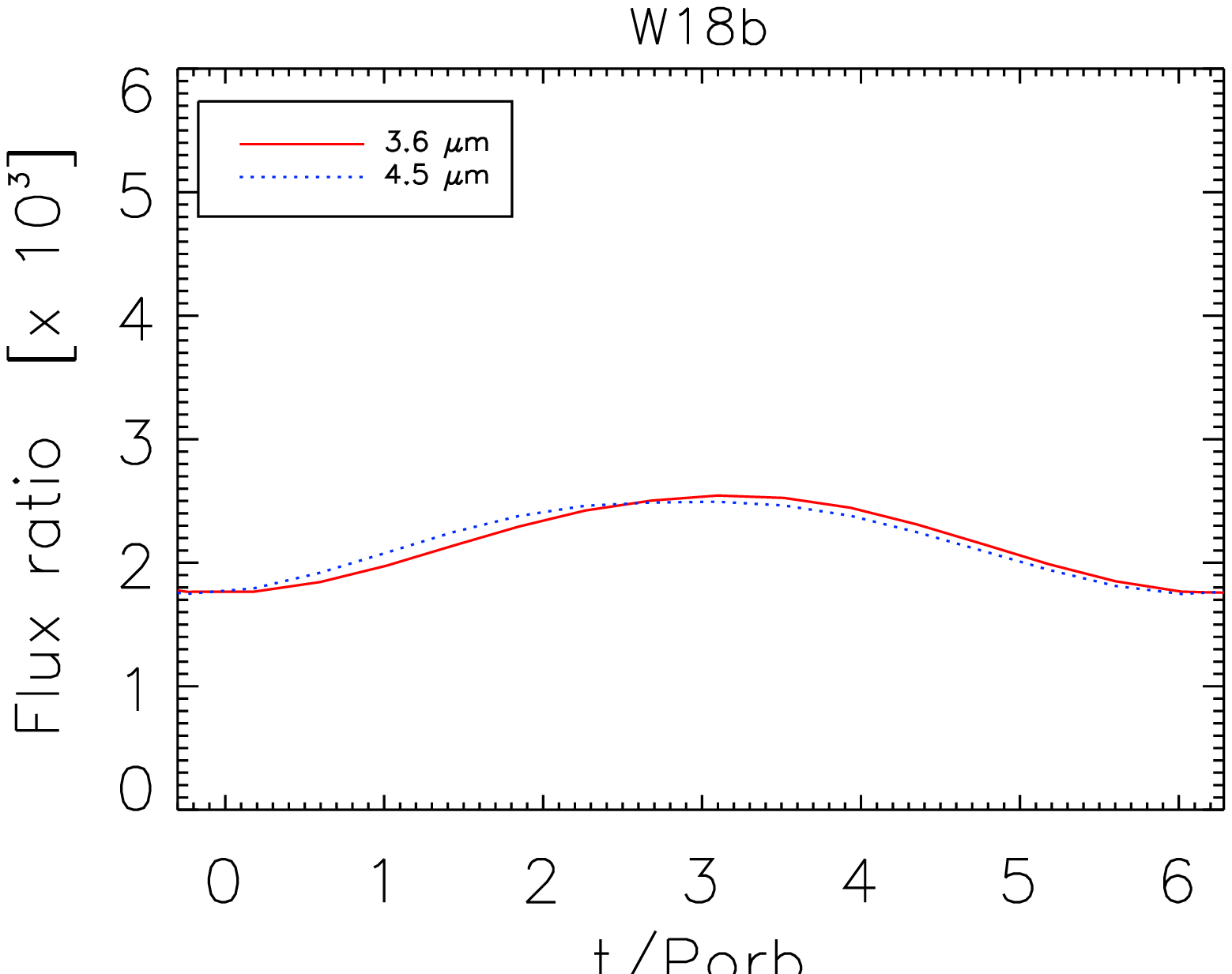}
    \caption{Flux ratio of the planet over the star\label{fig:curves} integrated in two Spitzer bandpass as a function of time for the two cases considered. $t = 0$ is when the substellar point is facing Earth. Upper panel: no rotation, lower panel: fast rotation.}
  \end{center}
\end{figure}

This planet is very well suited for measuring its light curve during a full orbit in order to constrain its heat redistribution by having the thermal curve as a function of the orbital  phase, in a similar way to \citet{Knutson07}.
Figure~\ref{fig:curves} show such phase curves for the two scenarios considered.
For a non rotating planet, as expected the maximum of flux occurs when the substellar point is facing the Earth ($t = 0$ in Figure~\ref{fig:curves}).
On the contrary, due to the radiative timescale of the atmosphere when a rotation is introduced, the maximum of flux is shifted and we measure a delay.
In the particular case presented here, since the period of rotation is about half the orbital period, the shift is about 180$^\circ$.

\section{Conclusions}
\label{Conclusions}

We presented a model for the planet WASP-18b in which the stellar heating has been modulated in order to account for the different longitude of the planet, with and without rotation.

In the case with a fast rotation, the temperature mixing is too efficient to match the observational data of secondary eclipse from \citet{Nymeyer10}.
The day side is too cool indeed to get the observed amplitude.

The no-rotation case, corresponding to a nill heat redistribution (in the longitudinal direction) fits very well the data of \citet{Nymeyer10}.
In that case, we expect a full phase curve to exhibit a large amplitude and no phase shift with respect to the substellar point.

This work does not pretend to investigate all the dynamical processes taking place in the  exoplanets atmospheres.
In particular, it neglects any latitudinal heat transport.
On the contrary, it aims at showing that a zonal circulation would be in this case very low.
Moreover, adding latitudinal transport would lead to a more uniform global structure. Given the agreement between our model without any transport at all  and the data from \citet{Nymeyer10}, this transport should also be minimal.
As reviewed in \citet{ShowmanPolvani}, in most 3D General Circulation Models, a zonal superrotating jet at the equator is the main feature that develops and can be observed in hot exoplanets. In that case our simpler approach can be representative of this feature.

A Spitzer thermal phase curve during a full orbit has already been measured for this planet \citep{Maxted12}. 
The data pertaining to the 3.6 and 4.5 microns light curves agree well with the model we present without rotation.
Detailed comparison of the model with the observed light curve will enable us to translate the rotation (here seen as sign of heat redistribution) into an upper limit of the physical wind speed of the atmosphere Ð slow for WASP-18b.
Moreover, an analysis of the shape and phase of the peak of the light curve will allow us to constrain the heat redeposition processes.

%A \emph{Spitzer} thermal phase curve has already been measured for this planet (GO program ID 60185).
%The data are currently under analysis and interpretation for the model to match the observed phase and amplitude is being performed, as well as a detailed %analysis of the influence of the chemical composition such as departure from the chemical equilibrium.
%Comparison of the model with the observed light curve will enable us to translate the rotation (here seen as sign of heat redistribution) into a\changes{n upper limit of the} physical wind speed of the atmosphere -- slow for \W18b .
%Moreover, an analysis of the shape and phase of the peak of the light curve will allow us to constrain the heat redeposition processes.

%% The Appendices part is started with the command \appendix;
%% appendix sections are then done as normal sections
%% \appendix

%% \section{}
%% \label{}

%% References
%%
%% Following citation commands can be used in the body text:
%% Usage of \cite is as follows:
%%   \cite{key}         ==>>  [#]
%%   \cite[chap. 2]{key} ==>> [#, chap. 2]
%%

%% References with bibTeX database:

%\bibliographystyle{elsarticle-num}
%\bibliography{<your-bib-database>}

%% Authors are advised to submit their bibtex database files. They are
%% requested to list a bibtex style file in the manuscript if they do
%% not want to use elsarticle-num.bst.

%% References without bibTeX database:

\end{document}